\documentclass[12pt, draftclsnofoot, onecolumn]{IEEEtran}
\ifCLASSINFOpdf
\usepackage[pdftex]{graphicx}
\else
\fi
\usepackage{silence}
\usepackage{amsmath,amssymb,amsfonts}
\usepackage{mathtools}
\usepackage{subcaption}
\usepackage{float}
\usepackage{comment}
\usepackage{placeins}
\usepackage{cite}
\usepackage{xcolor}
\usepackage{adjustbox}
\usepackage{multirow}
\usepackage{hyperref}
\usepackage{array}
\usepackage{graphicx}
\usepackage{textcomp}
\usepackage{mathtools}
\usepackage{breqn}
\usepackage{array}
\usepackage{dsfont}
\usepackage{mathrsfs}
\usepackage{braket}
\usepackage{pbox}
\newcolumntype{P}[1]{>{\raggedright\arraybackslash}p{#1}}
\ifCLASSINFOpdf
\else
\fi
\begin{document}
\title{Implementation of the Digital QS-SVM-based Beamformer on an FPGA Platform}
\author{Somayeh~Komeylian, and~Christopher~Paolini,~\IEEEmembership{IEEE,~Member}}
\markboth{}%
{Shell \MakeLowercase{\textit{et al.}}: Bare Demo of IEEEtran.cls for IEEE Journals}
\maketitle
\begin{abstract}
To address practical challenges in establishing and maintaining robust wireless connectivity such as multi-path effects, low latency, size reduction, and high data rate, the digital beamformer is performed by the hybrid antenna array at the frequency of operation of 10 GHz. 
The proposed digital beamformer, as a spatial filter, is capable of performing Direction of Arrival (DOA) estimation and beamforming. 
The most well-established machine learning technique of support vector machine (SVM) for the DoA estimation is limited to problems with linearly-separable datasets.

To overcome the aforementioned constraint, in the proposed beamformer, the QS-SVM classifier with a small regularizer has been used for the DoA estimation in addition to the two beamforming techniques of LCMV and MVDR. 
The QS-SVM-based beamformer has been deployed in an FPGA board, as demonstrated in detail in this work. 
The implementation results have verified the strong performance of the QS-SVM-based beamformer in suppressing undesired signals, deep nulls with powers less than -10 dB in undesired signals, and transferring desired signals. 
Furthermore, we have demonstrated that the performance of the QS-SVM-based beamformer consists of other advantages of average latency time in the order of milliseconds, performance efficiency of more than 90\%, and throughput of about 100\%.  

\end{abstract}
\IEEEpeerreviewmaketitle
\section{Introduction}
\IEEEPARstart{A}{daptive} digital beamformers have recently received numerous attention in a variety of applications such as MIMO 
wireless communications~\cite{Ioushua,Bogale,Sohrabi}, 
fault detection and calibration~\cite{Fulton,Ng,Singh2019}, 
wireless localization~\cite{Nallabolu,Cremer,Chen}, and 
mmWave communication systems~\cite{Hsu,Gimenez,Yu}. 

The involved problems of the framework of the antenna array signal processing have mainly included techniques of beamforming and DoA estimation. 
Techniques of DoA estimation are performed to obtain DoAs of sources, while beamforming algorithms are employed for tracking the desired signals in the real environment and nulled out other interference signals. 
In other words, techniques of DoA estimation provide a non-linear mapping from original signal sources to measured results in outputs of array elements, while beamforming techniques have focused on recovering the original signals from interested sources. 

In this work, we have rigorously implemented the quadratic surface support vector machine (QS-SVM) method for DoA estimation, in addition to the two beamforming techniques of minimum variance distortionless response (MVDR) and linearly constrained minimum variance (LCMV). 
The MVDR technique refers to a well-established adaptive beamforming algorithm to determine weight vectors for beamsteering. 
However, the MVDR technique shows a weak nullsteering performance on the interference signals. 
To overcome the aforementioned constraint, we have performed the LCMV technique in the preliminary stage of the proposed digital beamformer to suppress the interference signals. 
Therefore, the LCMV beamformer is capable of providing a higher directivity and lower sidelobe level. 

The DoA estimation has been fulfilled based on the QS-SVM method with a small regularizer. 
The practical implications of kernel-based classifier techniques, such as the SVM methods, are mostly limited only to supporting linearly-separable datasets. 
However, almost all practical problems have inherent non-linear characteristics.
To address this limitation, feasible methods for practical applications have to be capable of supporting non-linearly separable datasets without performing a mapping to a larger feature space. 
In other words, we are interested in developing feasible techniques, which include non-linear classifiers in the original spaces of their own datasets. 
The QS-SVM model~\cite{Boser} employs quadratic surfaces for classifying a non-linear dataset in its original space without mapping it to a higher dimensional feature space. 
Although the highly non-linear classification generated by the Gaussian kernel function or quadratic surfaces performs more effectively than the SVM techniques based on hyperplanes in terms of classification accuracy, we can still improve the performance of SVM techniques by producing hyperplane classifiers in the conventional QS-SVM models. 
The QS-SVM technique does not produce any classifier hyperplane for capturing linearly-separable datasets. 
To overcome this limitation, we have added a regularizer into the objective function of the QS-SVM model results, as discussed in detail in~\cite{Boser,LUO201889,LUO20201008,Luo,MALDONADO2017656,meyer01}.
The presence of hyperplanes yields an increase in the number of outliers in the training datasets, and thereby the QS-SVM model outperforms other models in terms of the generalization capability and robustness to outliers in addition to capturing linearly-separable datasets. 
A relatively large value of the penalty parameter of the regularizer can capture linearly-separable datasets. 
Furthermore, an appropriate choice of a small value of the penalty parameter of the regularizer results in more quadratic surfaces.

Maximizing SINR values is accompanied by not only an increase in the gain in the directions of desired signals, but also an enhancement in the spatial distribution of a radiation pattern of the antenna array. 
Hence, to further improve the beamforming performance, the antenna array has to be capable of having the horizontal and vertical spatial distribution of the radiation pattern to obtain an additional degree of freedom for suppressing interference. 
Indeed, beamforming techniques have been used as a spatial filter for reception and transmission by the antenna array. 
As described in details in~\cite{https://doi.org/10.48550/arxiv.2210.00317}, the hybrid antenna has demonstrated superior advantages over other available antenna arrays for massive wireless communications~\cite{https://doi.org/10.48550/arxiv.2210.00317}.

The proposed techniques for beamforming and DoA estimation are strongly capable of performing in real-time on hardware implementations. 
In this work, we have provided the implementation setup of the digital beamforming on an FPGA platform when performing the QS-SVM modeling of the proposed hybrid antenna array with bowtie elements. 

FPGA technology has recently received tremendous attention as it offers several advantages over  other available embedded processors, such as digital signal processing (DSP) and specific application integrated circuit (ASIC) devices~\cite{Hussain2018,Abusultan,Bilel,Kamali}:
\begin{itemize}
    \item Recent FPGA technology has programmable logic and the capability of algorithm parallelization for further enhancing power consumption, flexibility, and accuracy.

    \item Recent advances in the FPGA architectures include a higher storage density, a drastic reduction in power consumption and cost, a large number of gates, and a high-performance processor.

    \item Recent FPGA software and high-level optimizations have to be accompanied by architectural changes in the FPGA board in order to satisfy drastic computations of SVM-based applications. Advances in FPGA technology have rigorously presented high-level software tools to be easily adjusted to the FPGA hardware.
\end{itemize}

Furthermore, FPGAs have evolved over the years into more heterogeneous devices integrated with various types of "hard IP" blocks, such as PCIe solid-state drives (SSDs), floating-point DSPs, etc. Hard blocks can also be implemented in FPGAs to address specific applications. 

The deployment set-up of the proposed digital beamformer on the FPGA board consists of the software and hardware implementations. 
The software implementation involves the LCMV beamformer technique for performing nullsteering of the hybrid antenna array in a real environment, while the hardware deployment involves implementing a quadrature programming solver for the QS-SVM technique with a regularizer, weight updating using the MVDR technique, and inner multiplications of the QS-SVM on the FPGA platform. 

Consequently, in addition to the introduction section, this work is organized into the four sections: 
Section II is allocated for reviewing the related technical literature. 
The methodology and theoretical framework of the QS-SVM-based digital beamformer have been provided in section III. 
Section IV has focused on describing the deployment set-up on the FPGA platform, which includes software and hardware implementations.
Section~\ref{sec:performance} presents an evaluation of performance of the proposed digital beamformer, in term of throughput, latency, and efficiency. 
The conclusions of this work is presented in section~\ref{sec:conclusion}.

\section{Literature Review on the Related Work}

Digital beamforming deployment is significantly affected by the hardware platform, hardware design, system architecture, beamformer, and optimization techniques. 
Hence, the discussion in this section presents a brief review of the technical literature on implementing digital beamforming on FPGA accelerated hardware platforms.  

Dick \textit{et al.} in~\cite{Dick} have proposed real-time QR decomposition (QRD)-based beamforming on an FPGA platform. 
However, the significant disadvantage of the proposed beamforming approach consists in determining weights using the MVDR technique, without performing any nullsteering method. 
The MVDR technique is capable of performing beamsteering by determining the weight vectors; however, its nullsteering performance remains unsatisfactory. 
In this sense, when a mismatch occurs between the direction of the steered mainbeam and the direction of the signal of interest, the MVDR beamformer considers the reference signal as an interference signal and thereby strongly dissipates the signal of interest. 
Hence, in wireless communications with an inherent characteristic of strong multi-path effects, we have to perform a nullsteering beamformer in addition to determining weights such as LCMV beamformer technique. 
Furthermore, in~\cite{Dick}, the QRD algorithm has been employed for computing weights on an FPGA platform; however, it may not overcome the challenging requirements due to the fact that wireless communications involve computationally intensive operations on large volumes of data. 
Therefore, we have to deploy different strong optimization methods and beamformer techniques on a hardware platform to achieve satisfactory performance.  

FPGA implementation of adaptive digital beamforming for a massive array based on the conventional least mean squared (LMS) method has been proposed in~\cite{Lopes}. 
A major disadvantage of the proposed adaptive digital beamforming method is a dependency on the convergence speed of the conventional LMS on a spread of eigenvalues of the correlation matrix of the input signal. 
In other words, when the input signal consists of disparate eigenvalues, slow modes of convergence are dominant. 
This negative effect worsens the adaptive digital beamforming performance by increasing the number of antennas~\cite{Lopes}.  
Although the adaptive digital beamforming performance in terms of throughput, latency, and energy consumption has significantly enhanced, the strong mutual coupling effect between 64 antennas yields a performance degradation in terms of antenna efficiency, which is not reported in~\cite{Lopes}. 
Moreover, the spatial filtering functionality of the digital beamforming and its performance evaluation have not been represented in~\cite{Lopes}.     

The Cholesky decomposition or factorization has been performed on the input covariance matrix and results have shown a good performance under the computational complexity~\cite{Lu}. 
However, the obtained results and efficiency of Cholesky decomposition differ considerably in terms of the implementation and architecture details of the computing hardware, as reported in table 2 of~\cite{Lu}. 
Furthermore, the rectangular antenna array in~\cite{Lu} is very sensitive to the mutual coupling effect compared to a circular antenna array.

Xin \textit{et al.} in~\cite{Xin} implemented digital beamforming on an FPGA with a DSP processor. 
They have employed the Wiener vector solution for estimating the minimum mean squared error by minimizing a time-averaged squared error function.
Although the Wiener vector solution results in a drastic noise reduction, the involved solutions include computing partial derivatives with respect to both the real and imaginary parts, and requires both parts to be zero. 
Hence, its hardware implementation involves computational complexity, high operational costs, and low computational speed, which is not of practical interest in wireless communication channels.    

Ullah \textit{et al.} in~\cite{Ullah} proposed a millimeter-wave digital beamforming receiver on an FPGA platform to mitigate the interference between two adjacent channels in MIMO wireless communications. 
Although they could achieve superior performance in terms of maximum SIR values of approximately 36 dB in channel \#1, and 26 dB in channel \#2, the sensitivity of the hardware designs to the practical imperfections and losses in coaxial cables, the custom-designed encoder circuit board, and the adapter board presented a significant disadvantage.

\section{Methodology and Theoretical Framework}

Discussion in this section has focused on describing the theoretical and mathematical framework of the proposed QS-SVM-based digital beamformer. 
We aim to implement the hybrid antenna array with bowtie elements in~\cite{https://doi.org/10.48550/arxiv.2210.00317} for the proposed digital beamformer in this work, due to its superior performance in terms of a directive radiation pattern, the SINR values, and the antenna efficiency, as demonstrated in~\cite{https://doi.org/10.48550/arxiv.2210.00317}.   
The digital beamformer has to be capable of performing the QS-SVM optimization method for measured outputs of the hybrid antenna array with bowtie elements of Fig.1 and table 1, as demonstrated in Fig.2.
Furthermore, the proposed optimization methods should be capable of overcoming high computational complexities and transferring very large volumes of data in massive wireless communications.  

\begin{figure}[htbp]
\centering
\includegraphics[width=0.75\textwidth,height=0.45\textwidth]{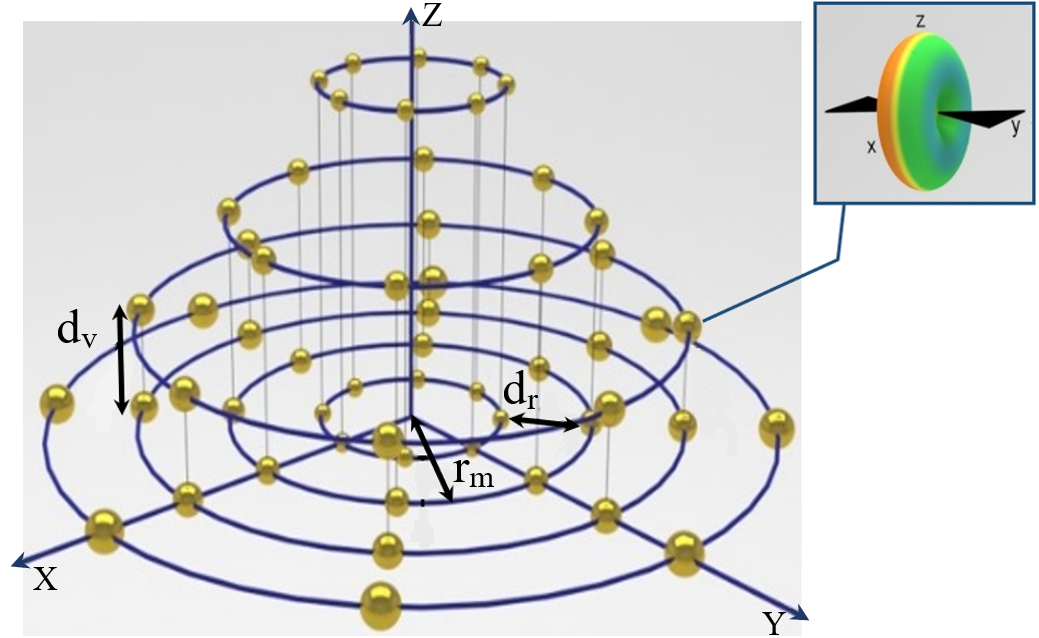} \\
(a) \\
\includegraphics[width=0.3\textwidth,height=0.4\textwidth]{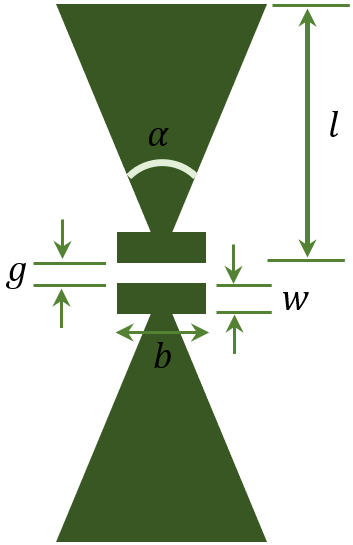} \\
(b) \\
\caption{Demonstration of the 3D configuration of the hybrid antenna array with bowtie elements at the frequency of the operation of 10 GHz with bowtie elements. The assigned design parameters of the hybrid antenna array are listed in detail in the Table 1. In this work, each hybrid antenna array consists of three cylindrical antenna arrays~\cite{Hussain}, and one circular antenna array~\cite{Noordin}. (b) A top view of the bowtie antenna; $l$: arm length, $\alpha$: flare angle, $g$: feed gap, $w$: bar width, $b$: bar length. In this work, the design parameters of each bowtie element are assumed to be $l$=6 mm, $\alpha$=60°, $g$=0.02 mm, $w$=0.02 mm, and its thickness is equal to 0.01 mm.}
\label{fig:configuration}
\end{figure}

\begin{table}[ht]
\centering
    \begin{tabular}{c|c|c}
Parameters &  Definition & Value \\ \hline
        $N_h$ & Number of elements of any circular loop &	$N_h=20$ \\
        $Q_h$ & Number of elements of any cylinder & $Q_h=40$ \\
        $M_h$ & Total number of cylinders in the proposed array & $M_h=3$ \\
        $P_h$ & Number of circular loops in the cylinder & $P_h=2$ \\
        $d_v$ & Vertical spacing between two consecutive circular loops & $d_v=0.5\lambda$ \\
        $d_r$ & Horizontal spacing between two consecutive circular loops &	$d_r=0.5\lambda$ \\
        $\phi,\theta$ & Maximum scanning angles & $\phi=45^{\circ}, \theta=45^{\circ}$
    \end{tabular}
    \caption{The design parameters for the proposed hybrid antenna array.}
    \label{tab:parameters}
\end{table}
We have demonstrated that the proposed methodology and techniques are well-matched for beamforming applications due to their simplicity to be implemented on the hardware platform as well as to overcome limitations of modern wireless communication channels~\cite{Xiao}. 
\begin{figure}[ht]
\centering
\includegraphics[width=0.8\textwidth, height=0.6
\textheight]{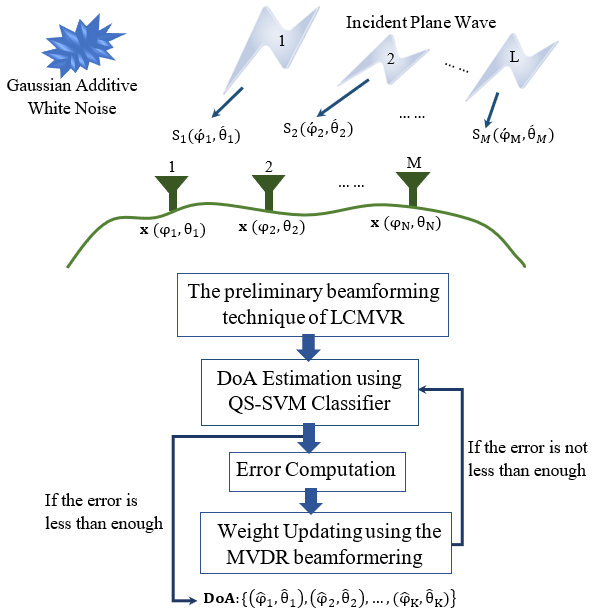}
\caption{Procedures for the DoA estimation using the QS-SVM-based digital beamformer. It is assumed that there are $M$ hybrid antenna array. Each hybrid antenna array consists of 57 elements whose outputs have been measured. Here, position vectors are supposed to be in a two-dimensional plane of $\theta$ and $\phi$.}
\label{fig:algorithm}
\end{figure}

\subsection{Methods of Modeling and Producing Data}

In the real environment, we assumed that the available sources emit sinusoidal signals. 
Data modeling is the process of producing data based on assumptions of the real environment to generate signal sources. 
In this study, we have employed the MATLAB programming platform to provide raw data. 
Since Mathworks functions of the MATLAB programming platform have been extensively tested, evaluated, and verified based on IEEE standards and criteria, our obtained simulation results through MATLAB provide a very high level of realistic accuracy. 
In the preliminary stage of producing data, a sinusoidal function is considered as an available signal. 
In this scenario, the available signal is significantly affected by environmental noise, the coupling effect, and communication channel conditions. 
In this study, conditions of the communication channel are supposed to be the Ricean fading channel, line of sight (LoS), and equal correlation matrices, as described in detail in~\cite{Tataria}.

In addition, the additive white Gaussian noise (AWGN) is added to the simulated signal by the standard Mathworks' function of AWGN. 
Afterward, the standard Mathworks function of \textit{collectPlaneWave()} exerts the plane wave condition to the simulated signal or incoming signal to the hybrid antenna array, which has experienced the AWGN noise and channel conditions.
In this work, the white Gaussian noise has a variance of $\sigma^2$. 
The noise model of $\mathbf{n}$ consists of the covariance matrix of $\sigma^2 \mathbf{I}$, which is a centered complex vector. 

\subsection{The Beamforming Model}

Signal measurements at the outputs of array elements of the hybrid antenna array are modeled by random vectors. 
Ultimately, a beamforming model consists of parameters, such as the source power and covariance of the noise power.  
As demonstrated in Fig.~\ref{fig:configuration}, an incident plane wave is expressed in terms of its generated source and locations in which measured. 


Beamforming is employed for calculating the scalar product between the measured data at outputs of array elements and the steering vector in the following equation, 
\begin{equation}
    | \braket{\mathscr{R}_n(\phi_{l},\theta_{l}),\mathbf{h}_n(\phi_{l},\theta_{l})}|^2
\end{equation}
\noindent for $l=1, 2,\ldots,L$
and $n=1, 2,\ldots,N$.
Eq.~\ref{eq:mean} represents the mean of the estimated power of the source,

\begin{equation} 
    \mathscr{R}_n(\phi,\theta) = \sum_{l = 1}^{L} \left(
     \underbrace{\mathbf{C}_n e^{j \mathbf{K}.\mathbf{r}_n}}_{\text{source}} + \underbrace{\mathbf{v}_n}_{\text{noise}}
    \right)
    \label{eq:mean}
\end{equation}

\noindent where $\mathbf{v}_n = \sigma^2 \mathbf{I}$ for $n = 1,2,\ldots,N$.

The steering vector of the proposed hybrid antenna array in Fig.~\ref{fig:configuration} is obtained by the multiplication of the steering vector matrices of the three cylindrical antenna arrays and circular antenna array in the following equations~\cite{Tan}, 

\begin{equation}
    h_n^{(T)}(\phi,\theta) = \underbrace{\frac{1}{N^{(1)}} \frac{\mathbf{C}_n^{(1)}(\phi,\theta)}{|\mathbf{C}_n^{(1)}(\phi,\theta)|} \otimes
    \frac{1}{N^{(2)}} \frac{\mathbf{C}_n^{(2)}(\phi,\theta)}{|\mathbf{C}_n^{(2)}({\phi,\theta})|}\otimes
    \frac{1}{N^{(3)}} \frac{\mathbf{C}_n^{(3)}({\phi,\theta})}{|\mathbf{C}_n^{(3)}({\phi,\theta})|}}_{\text{for the three cylindrical antenna arrays}}\otimes
    \underbrace{\frac{1}{N^{(4)}} \frac{\mathbf{C}_n^{(4)}({\phi,\theta})}{|\mathbf{C}_n^{(4)}({\phi,\theta})|}}_{\text{for the circular antenna array (when h=0)}}
\end{equation}

\noindent in which the two-dimensional steering vector of each cylindrical antenna array in terms of phase differences or coefficients of the source vector is given by,

\begin{equation}
    \textbf{h}_n (\phi,\theta) = \frac{1}{N} \frac{\mathbf{C}_n(\phi,\theta)}{|\mathbf{C}_n(\phi,\theta)|} 
\end{equation}

\noindent where coefficients of the source vector are expressed by,
\begin{equation}
\mathbf{C}(\phi,\theta) = 
\left[
\begin{matrix}
    g_1(\phi,\theta) e^{-\frac{j2\pi}{\lambda}(r \sin \theta \cos(\theta - \theta_1) + h \cos \phi)} \\
    g_2(\phi,\theta) e^{-\frac{j2\pi}{\lambda}(r \sin \theta \cos(\theta - \theta_2) + h \cos \phi)}\\
    \vdots \\
    g_N(\phi,\theta) e^{-\frac{j2\pi}{\lambda}(r \sin \theta \cos(\theta - \theta_N) + h \cos \phi)}
\end{matrix}
\right]
\end{equation}

\begin{equation}
    \mathbf{K} = \frac{2\pi}{\lambda}(K_x,K_y,K_z) = (\sin \phi \sin \theta, \sin \phi \cos \theta, \cos \phi)
\end{equation}

\noindent where $\mathbf{K}$ represents the wavelength vectors in directions of radius vectors of $\mathbf{r}_n$ pointed to $n$th array elements. Here, position vectors are assumed to be in a two-dimensional plane of $\theta$ and $\phi$.

\begin{equation}
    \mathbf{r}_n^T = [r \cos \theta_n, r \sin \theta_n, h]^T
\end{equation}
\noindent Hence, the phase shift relative to the origin is given by, 
\begin{equation}
    \mathbf{K}.\mathbf{r}_n = \frac{2\pi}{\lambda} [\sin \phi \sin \theta \hspace{4mm} \sin \phi \cos \theta \hspace{4mm} \cos \phi]
    \left[
    \begin{matrix}
    r \cos \theta_n \\ r \sin \theta_n \\ h
    \end{matrix}
    \right] = \frac{2\pi}{\lambda}(r \sin \phi \cos \theta \cos \theta_n + r \sin \phi \cos \theta \sin \theta_n + h \cos \phi)
\end{equation}
Furthermore, $g_n(\phi,\theta)$ refers to the antenna gain at the $n$th element.

\subsection{The DoA Estimation Model} 
In this section, we have provided the theoretical and mathematical procedures for the DoA estimation of a desired source using the kernel-free quadratic SVM method. 
The QS-SVM model is proposed for direct classification by the quadratic function, which is applicable for linear and non-linear datasets.  

\textbf{The training procedure:} The supervised learning algorithms are designed to construct a model within the training phase. 
The machine learning algorithm analyzes the training dataset to generate the parameters for classifying quadratic surfaces such that there is a maximum margin between the different classes of data. 
Therefore, the output is a model trained by the training dataset.

Suppose a non-linearly separable training dataset with correct values of outputs are in pair with correct values of inputs in the following form,
\begin{equation}
\mathcal{I} = \{x_i,y_i^c \}_{i=1, c=1}^{i=n, q=Q}
\label{eq:I}
\end{equation}
%
Eq.~\ref{eq:I} holds for the $n$ index of training pattern and the $Q$ number of classes.
The training label is defined in the following equation,
\begin{equation}
y = i\in \{1,2,...,G\}
\label{eq:y}
\end{equation}
\noindent where $G$ refers to a number of output labels.

During the training procedure, the QS-SVM algorithm not only generates the hyperplane classifiers but also updates parameters and hyperparameters $(\textbf{w}, \textbf{b}, c)$ for changing the locations of hyperplanes so that two classes with the largest margin can be separated,  
\begin{equation}
\frac{1}{2}(\mathbf{x}^i)^T \mathbf{w}\mathbf{x}^i + \mathbf{b}^T\mathbf{x}^i + c
\label{eq:quadrature-surface}
\end{equation}

\noindent where Eq.~\ref{eq:quadrature-surface} represents the quadrature surface with the largest margin, and $\textbf{w}$ and $c$ represent the weighting vector perpendicular to the quadrature surface, and the bias value for shifting the quadrature surface parallel to itself, respectively~\cite{Komeylian2021-2}. 
In other words, the proposed quadratic surface of Eq.~\ref{eq:quadrature-surface} is employed for separating $n$ training points of the given training dataset. 
 
The machine learning algorithm analyzes the training dataset and constructs quadratic surface classifiers from the training dataset such that the sum of the relative geometrical margins becomes maximum at all training points in which there is no training point between two-quadrature surfaces, as described in Eqs. from 12 to 16.  

\noindent By minimizing 
\begin{equation}
\min \sum_{i=1}^n \|\textbf{w} \mathds{x}^i + \textbf{b} \|_2^2 + \hat{\eta}\sum_{i=1}^n \xi_i
\end{equation}

\noindent Subject to constraints:
\begin{eqnarray}
y^i\left( \frac{1}{2} (\textbf{x}^i)^T \textbf{w} \textbf{x}^i + \textbf{b}^T\mathbf{x}^i + c\right) \ge 1 - \xi_i, 
\\\xi_i \ge 0, \hspace{0.5cm} i = 1,\ldots,n, \\
\textbf{w} = \textbf{w}^T \in \mathds{R}^{m\times m}, \\
(\textbf{b},c)  \in \mathds{R}^{m} \times \mathds{R}^{1}.
\end{eqnarray}
In a condition, in which the training dataset cannot be quadratically separated or $G(\textbf{x}) = 0$, training points should satisfy one of the following conditions:

\noindent \textbf{Condition \#1:}
$\textbf{x}^i, \textbf{y}^i = -1,$ but 
\begin{equation}
\frac{1}{2} (\textbf{x}^i)^T \textbf{w} \textbf{x}^i + \textbf{b}^T\textbf{x}^i + c > -1, \end{equation}

\noindent \textbf{Condition \#2:}
$\textbf{x}^i, \textbf{y}^i = +1,$ but
\begin{equation}
\frac{1}{2} (\textbf{x}^i)^T \textbf{w} \textbf{x}^i + \textbf{b}^T\textbf{x}^i + c < +1.
\end{equation}

\noindent \textbf{The testing procedure:} The testing dataset, or the remainder of the dataset which was not employed for the training dataset, is fed to the model. 
As soon as the testing dataset is fed to the model, the model becomes fixed such that it cannot change. 
Then, the quadratic surface is used to separate the testing dataset. 
In other words, to assess how well the model can process the real-world data and generate accurate predictions, we employ the unseen dataset, or testing dataset. 

In the mathematical sense, the decision function of $D_{ij}(x)$ can be expressed by, 

\begin{equation}
D_{ij}(\textbf{x}) = \frac{1}{2}(\textbf{x})^T \textbf{w}_{ij}\textbf{x} + \textbf{b}_{ij}\textbf{x} + c_{ij}
\end{equation}

\noindent where $D_{ij}(x)=-D_{ji}(x)$

Therefore, the quadratic surface is applied to decide which class each data point of the testing dataset belongs to. 

\begin{equation}
    D_i(f(\textbf{x})) = \sum_{i=1,j\ne i}^{n} sign(D_{ij}\left(\textbf{x})\right)
\end{equation}

\noindent where

\begin{equation}
  sign(\mathbf{x}) =
    \begin{cases}
      1 & \text{for  $x\ge 0$}\\
      -1 & \text{for $x < 0$}
    \end{cases}       
\end{equation}


\section{Implementation Setup of the QS-SVM-based Beamformer on the FPGA board}

\begin{figure*}
\begin{subfigure}{1.0\textwidth}
\centering
\includegraphics[width=1.0\linewidth]{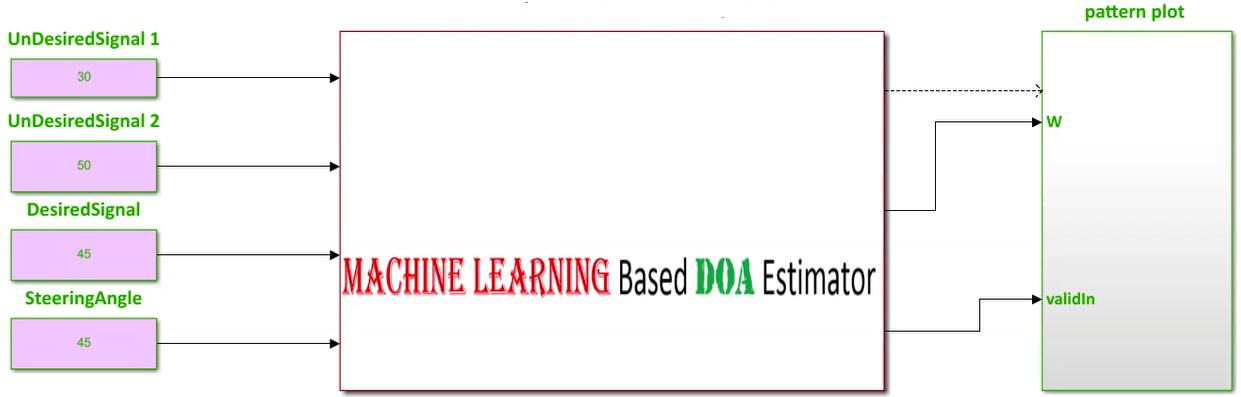}
\caption{} \label{fig:doa_Estimator_a}
\end{subfigure}
\begin{subfigure}{1.0\textwidth}
\centering
\includegraphics[width=1.0\linewidth]{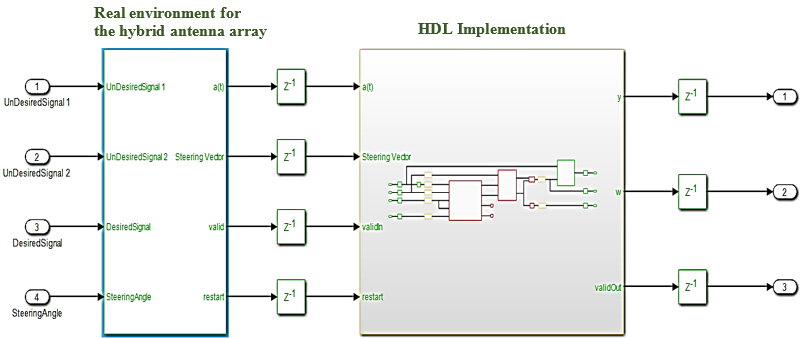}
\caption{} \label{fig:doa_Estimator_b}
 \end{subfigure}%
\caption{(a) The demonstration of the proposed QS-SVM-based digital beamformer. Here, position vectors are assumed to be in a plane of $\theta$ variable and $\phi$ constant.An angle of $45^\circ$ is the direction of a desired signal, which the steering angle should be identical to it. However, angles of $30^\circ$, and $50^\circ$ represent the directions of undesired signals, and (b) the demonstration of the proposed QS-SVM-based digital beamformer in the real and hardware environments.}
\label{fig:doa_Estimator}
\end{figure*}

The discussion in the previous sections has focused on providing the involved theoretical and mathematical techniques and solutions to implement the proposed QS-SVM beamformer. 
In this section we have represented the efficient and compact hardware implementations of the proposed approaches and solutions for the DoA estimation and beamforming techniques used in the QS-SVM beamformer. 
The technical approaches and solutions to deploy the proposed digital beamformer on the FPGA platform have been employed for two different environments: the real environment and the hardware environment, as demonstrated in Fig.~\ref{fig:doa_Estimator_b}.

\subsection{Real Environment and Software Implementation}

Incident signals impinging on elements of the hybrid antenna should be magnified and down converted to any convenient intermediate range of frequency. 
Phase references, or here phases of desired signals, remain unchanged for all operations of front ends. 
Then, the signal output of any front end should be digitalized by the $A/D$ converter. 

The performance of the proposed QS-SVM-based beamformer would be further clarified by performing the FPGA-based implementation for a practical example in the real environment, Fig.\ref{fig:doa_Estimator_a}.  
It is assumed that the direction of the desired source is $45^\circ$, and three signals in directions of $45^\circ$, $30^\circ$, and $50^\circ$, have impinged on the proposed hybrid antenna array. 
The preliminary stage of the spatial beamforming is to obtain the steering vector of the desired source, as described in detail in the previous sections. 
Therefore, we have computed the steering vector of the desired source in a direction of $45^\circ$. 
Only one of three signals, which impinged on the hybrid antenna array, is aligned with the direction of the desired source. 
In this sense, the adaptive beamforming technique of LCMV can significantly cancel interference signals and strongly steer the mainbeam of the radiation pattern of the proposed antenna array towards the direction of the desired source. 


\begin{figure*}[ht]
\begin{adjustbox}{addcode={\begin{minipage}{\width}}
{\caption{The spatial beamforming technique in the real environment. In this scenario, three signals have impinged on the hybrid antenna array and the steering vector for the desired source has been obtained at a direction of $45^\circ$.}
\label{fig:realenv}
\end{minipage}},rotate=90,center}
\includegraphics[width=1.3\textwidth]{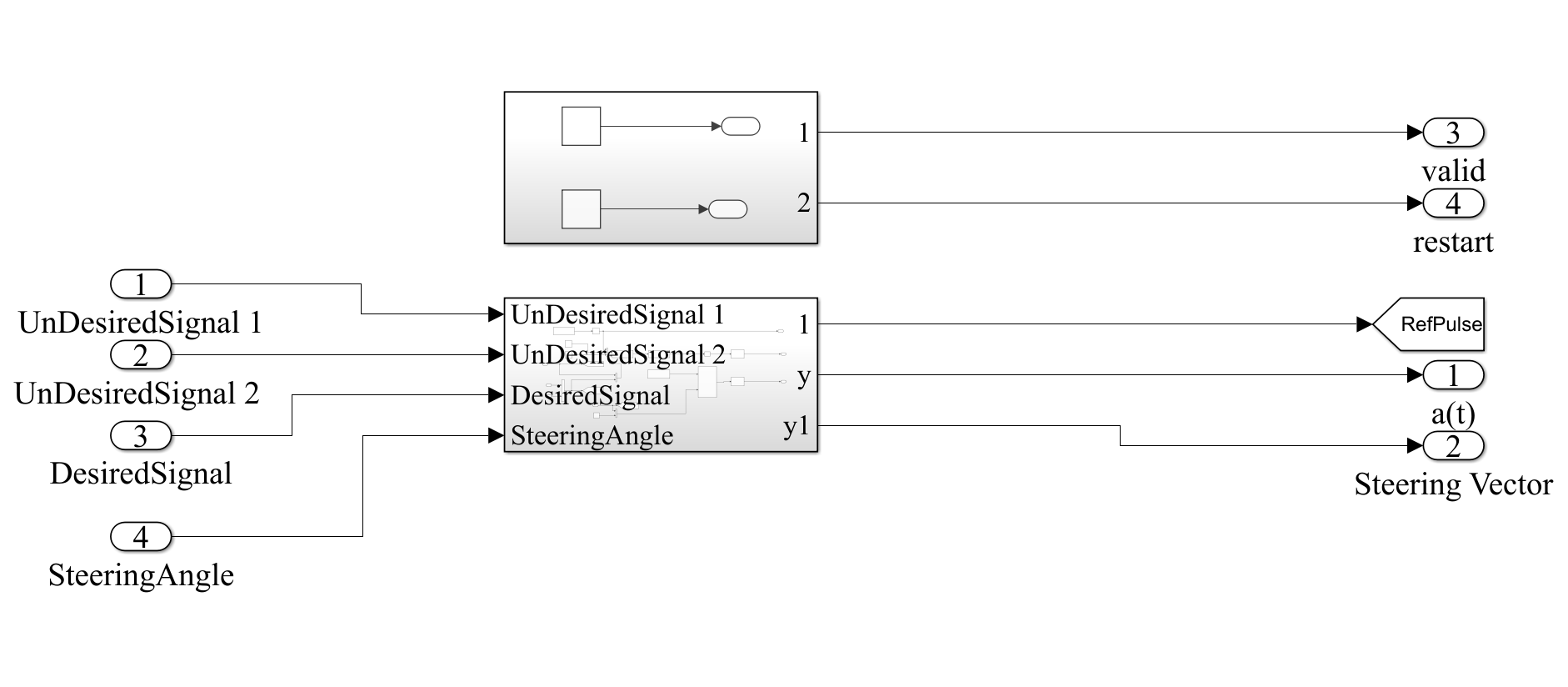}
\end{adjustbox}
\end{figure*}

\begin{figure*}
\begin{adjustbox}{addcode={\begin{minipage}{\width}}
{\caption{The aforementioned subsystem in Fig.~\ref{fig:realenv},
Beamforming technique of LCMV and other mathematical operations in the real environment. Simulink demonstrations of all mathematical operations in the MATLAB platform with detail.}\end{minipage}},rotate=90,center}
\includegraphics[width=1.35\textwidth]{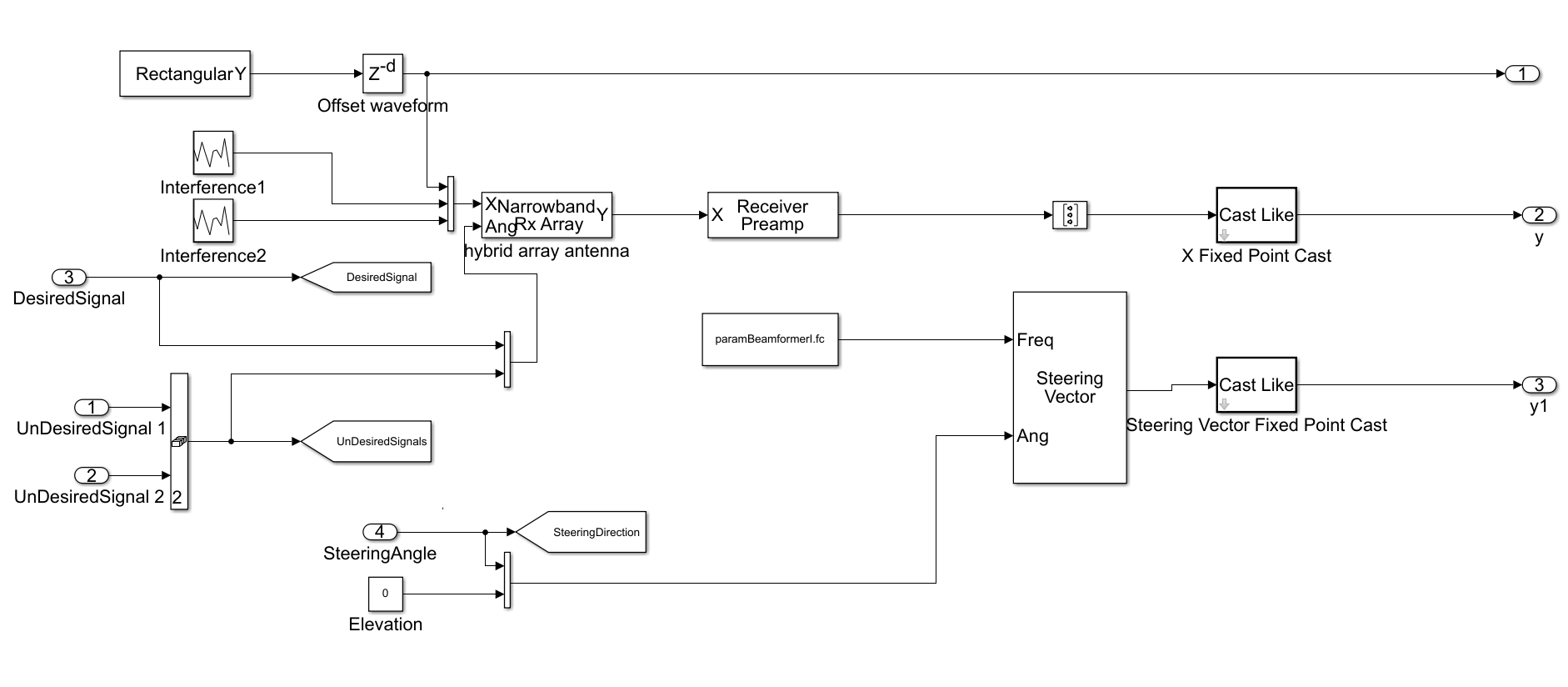}
\end{adjustbox}
\label{fig:simulink}
\end{figure*}

Eqs. from 11 to 19 have demonstrated mathematical operations and spatial constraints involved with the proposed hybrid antenna array on the FPGA board.
Angles of impinging signals have to be adaptively computed by the preliminary beamforming technique in the real environment. 
Beamforming techniques based on recognizing spatial locations of signals are substantially degraded by imperfections of array elements, mutual coupling, pointing errors, and multi-path effects in wireless communications. 
Especially, the LCMV algorithm may fail to perform appropriately when desired signals are expanded in all directions due to multi-path effects. 

\subsection{Hardware Environment and FPGA Implementation}

The proposed beamformer technique for the real environment is very sensitive to multi-path effects, which is an inherent property of wireless communication channels. 
Therefore, in this section, a further enhancement in the performance of the QS-SVM-based beamformer has been achieved by the hardware implementation. 
All MATLAB codes of the proposed QS-SVM-based beamformer, including data collection codes, LCMV and MVDR beamforming codes, the proposed hybrid antenna array codes, and codes of the QS-SVM optimization method, have to be efficiently programmed into a hardware description language (HDL) in the format of arithmetic operations, such as multiplications and/or additions.
Three main parts of the hardware implementation of the QS-SVM-based digital beamformer on the FPGA platform, including quadrature programming solver, weight updating, and SVM inner multiplications, have been rigorously demonstrated in Figs. from 3 to 10.
\begin{figure*}
\begin{adjustbox}{addcode={\begin{minipage}{\width}}
{\caption{Hardware implementation of the proposed digital SVM-based beamformer on the FPGA platform.}\end{minipage}},rotate=90,center}
\includegraphics[height=12.5cm, width=1.3\textwidth]{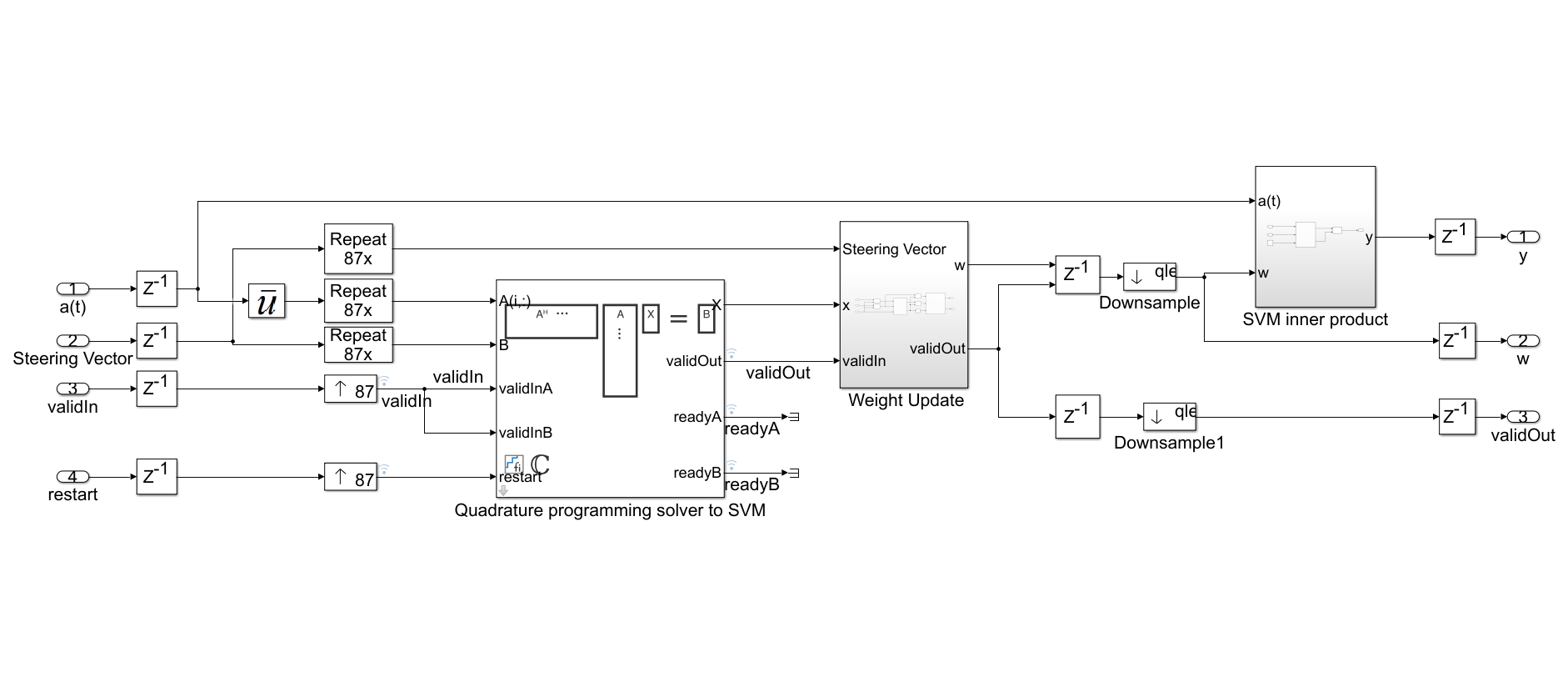}
\end{adjustbox}
\label{fig:implementation}
\end{figure*}

\textbf{Quadrature Programming Solver:}
The SVM algorithm allows for estimating a function, which maps the input data to a finite set of output labels~\cite{Boser,LUO201889,LUO20201008,Luo,MALDONADO2017656,meyer01} as discussed in detail in the previous section. 
Hence, the QS-SVM-based beamformer can overcome all constraints of the spatial reference techniques and achieve a superior performance in terms of SINR. 
There is no information about the probability distribution of the input data. 
Therefore, we have to minimize errors between the array output and the reference signals (or here desired signals) for the number of available data in the left hand sight of Eq.~\ref{eq:quadrature}, \begin{equation}
    \mathcal{R}(f) = \int_{\Omega} L(f(\textbf{x}_i),\textbf{y}_i) \hspace{0.1cm}dP(\textbf{x}_i,\textbf{y}_i) \approx \sum_{i=1}^N \textbf{w}_i \hspace{0.1cm}L(f(\textbf{x}_i),\textbf{y}_i)
    \label{eq:quadrature}
\end{equation}

\noindent where $L$ refers to the loss function. 

\begin{figure*}
\centering
\includegraphics[width=\textwidth]{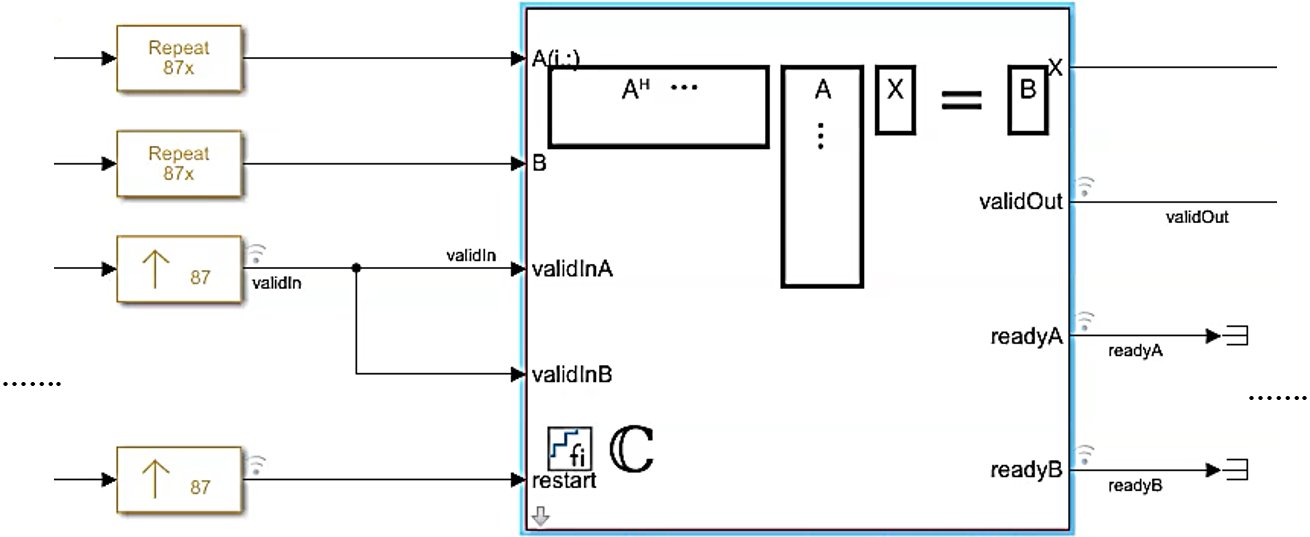}
\caption{Quadrature programming solver for the SVM algorithm in the HDL implementation}
\label{fig:quadrature}
\end{figure*}

\begin{figure*}
\begin{adjustbox}{addcode={\begin{minipage}{\width}}
{\caption{The demonstration of the complex Q-less QR forward backward substitute on the FPGA board.}
\label{fig:qr}
\end{minipage}},rotate=90,center}
\includegraphics[height=16cm, width=1.3\textwidth]{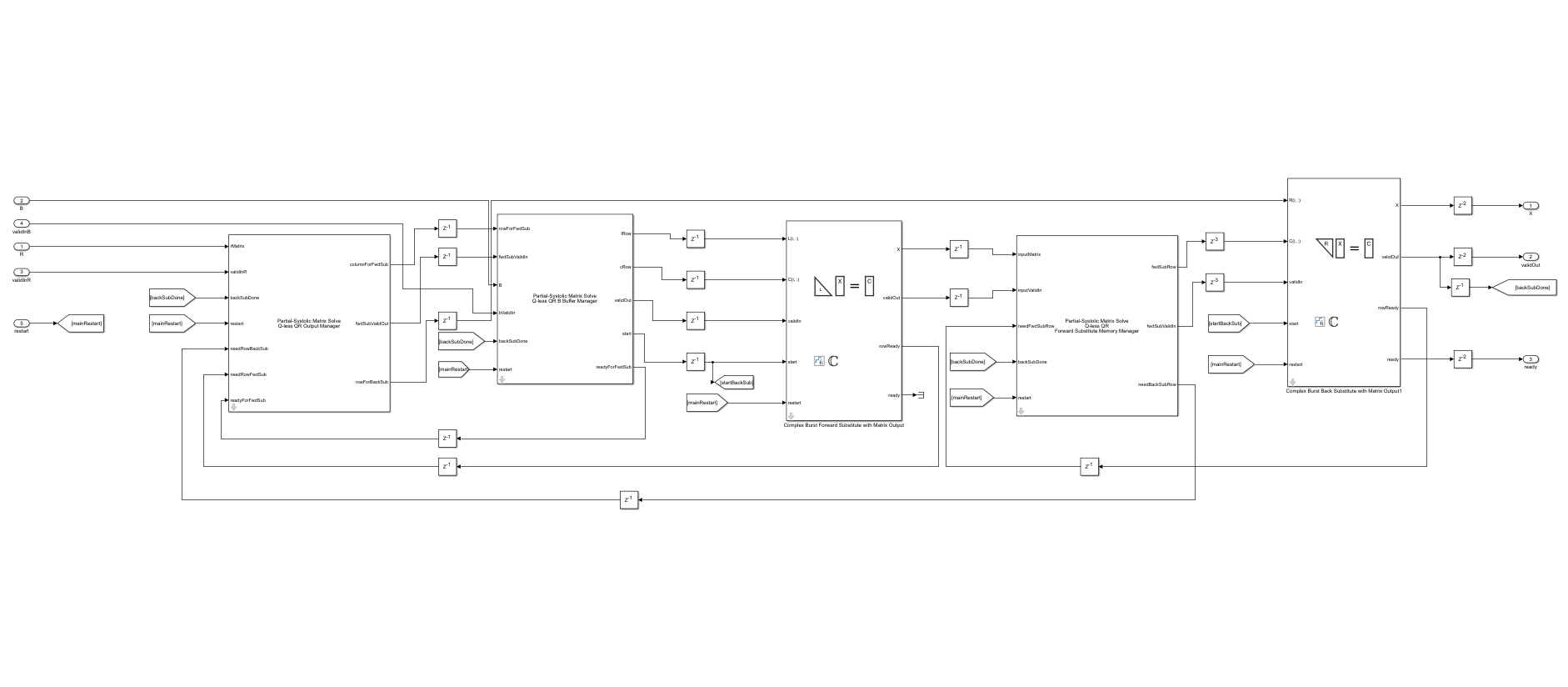}
\end{adjustbox}
\end{figure*}

We have employed the approach of Q-less QR decomposition with a forgetting factor to deploy the QS-SVM-based beamformer on the FPGA hardware.
In the preliminary stage, the integral of a continuous function, based on quadrature programming, has been expended in terms of a superposition of the weighted loss function at available data points, as represented in the right-hand sight of Eq.~\ref{eq:quadrature}. 
As demonstrated in Fig.~\ref{fig:qr}, we have rigorously produced $f(\textbf{x}_{i})$ using the approach of Q-less QR decomposition, as described in detail~\cite{LUO201889,LUO20201008,Luo,MALDONADO2017656,meyer01}. 
When an upper triangular factor becomes ready, we have computed the loss function within the forward and backward substitution. 
The upper triangular matrix is scaled by the square root of the forgetting factor. 
Based on the obtained value of the loss function, we have updated weights using the weight updating blocks. 

\textbf{Weight Updating:}
Parameters of a wireless communication channel have to vary quickly, thereby weight updating should be performed at a higher rate than a statistic scenario.
An alternation solution for updating the weight vector consists of directly computing the inverse of the correlation matrix using the MVDR algorithm. 
The primary goal of the MVDR beamforming technique is to minimize SINR expressed by the following formula~\cite{Darzi,Vorobyov},
\begin{equation}
    \text{SINR} \triangleq \frac{{| \braket{(\mathbf{C}_k e^{j \mathbf{K}.\mathbf{r}_k}),(\mathbf{h}(\phi_{k},\theta_{k})}|^2}}
    {{| \braket{(\mathbf{C}_k e^{j \mathbf{K}.\mathbf{r}_k} + \mathbf{v}_k +\mathbf{i}_k),(\mathbf{h}(\phi_{k},\theta_{k})}|^2}}
    \label{eq:sinr}
\end{equation}

\noindent Hence, the minimum SINR is achieved by the MVDR beamforming technique when the following conditions are satisfied,

\begin{align}
    & \min_\alpha \braket{(\mathbf{C}_k e^{j \mathbf{K}.\mathbf{r}_k} + \mathbf{v}_k +\mathbf{i}_k),(\mathbf{h}(\phi_{k},\theta_{k})}\\
    & \text{ s.t.} \\
    & \hspace{7mm} (\mathbf{h}^H(\phi_{k},\theta_{k},
    \textbf{C}_k(\phi,\theta)=1
\end{align}
\noindent Hence, the MVDR beamforming solution is expressed,
\begin{equation}
    \textbf{w}_{(MVDR)} =\alpha\mathbf{h}^{-1}({\phi_{-n+k},\theta_{-n+k})}|^2 \textbf{C}_k(\phi,\theta)
    \label{eq:MVDR}
\end{equation}

\noindent The normalized constant of $\alpha$ in Eq.~\ref{eq:MVDR}, which does not have any effect on the SINR, can be omitted.
\begin{equation}
\alpha = \frac{1}{{| \braket{(\mathbf{C}_k),(\mathbf{h}^{-1}({\phi_{-n+k},\theta_{-n+k}}))}|^2}} 
\end{equation}
%
  





\textbf{SVM Inner Product:}
Hadamard multiplication (or element-wise multiplication) refers to the multiplication of array objects, which includes both of the Hadamard product of $a*b$ and the matrix product of $a\hspace{0.5mm}@\hspace{0.5mm}b$, as demonstrated in Figs. 9 and 10. 

\begin{figure*}
\centering
\includegraphics[width=1.0\textwidth]{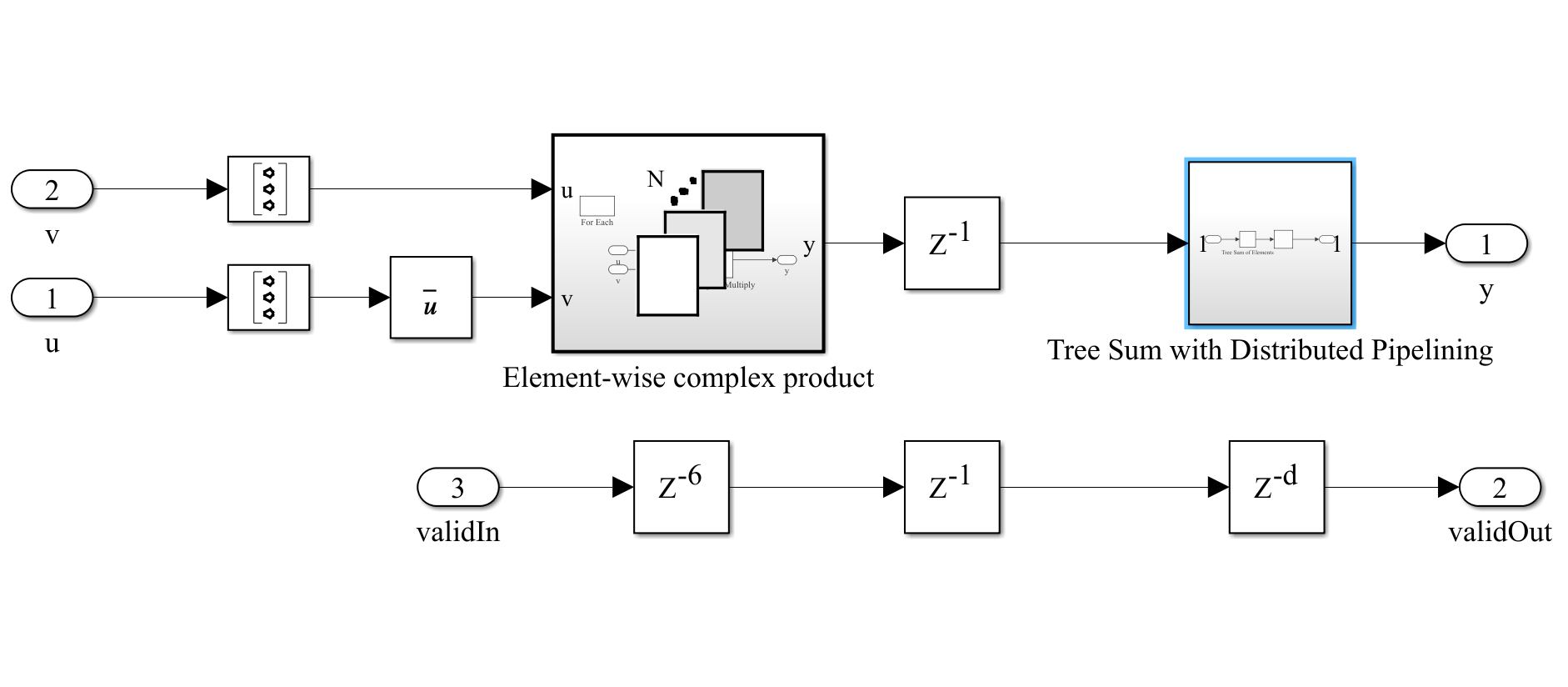}
(a)
\includegraphics[width=\textwidth]{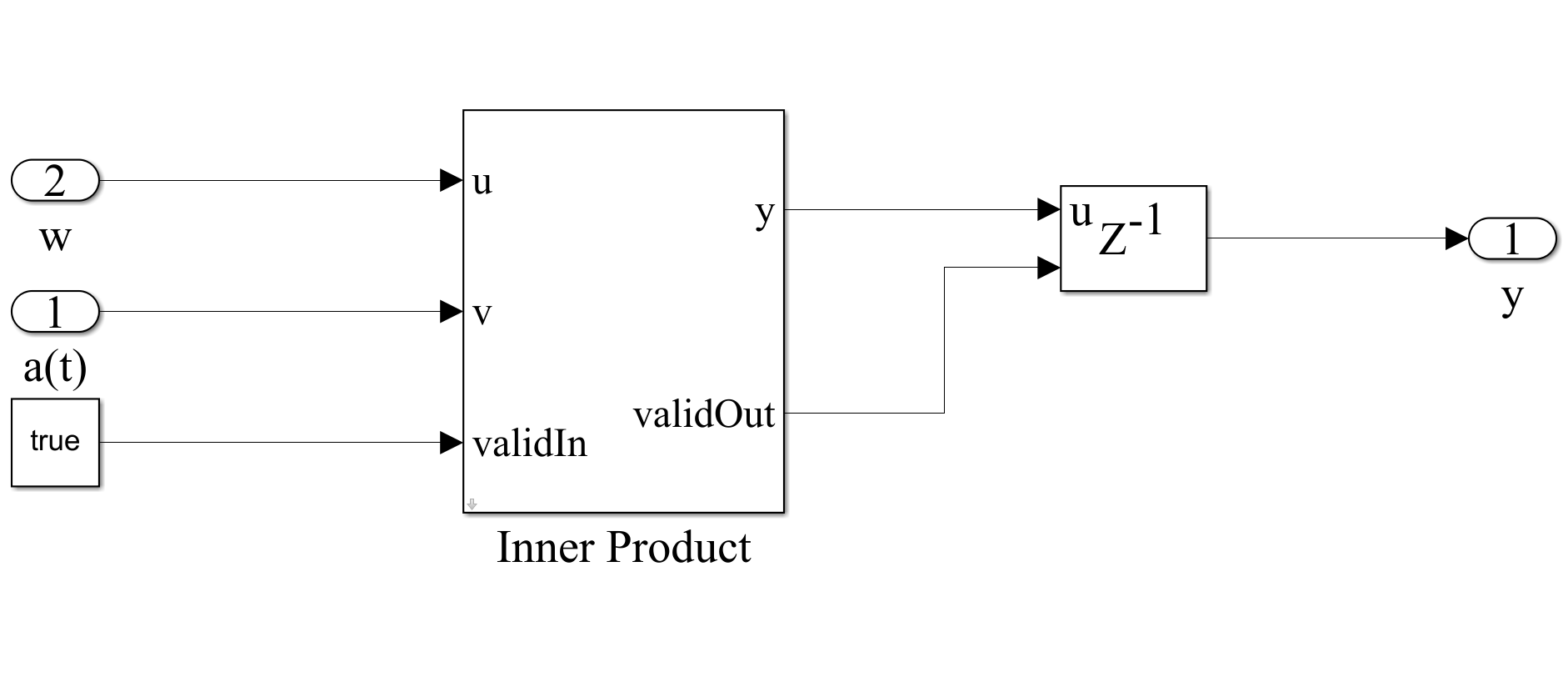}
(b) \\
\includegraphics[width=0.45\textwidth]{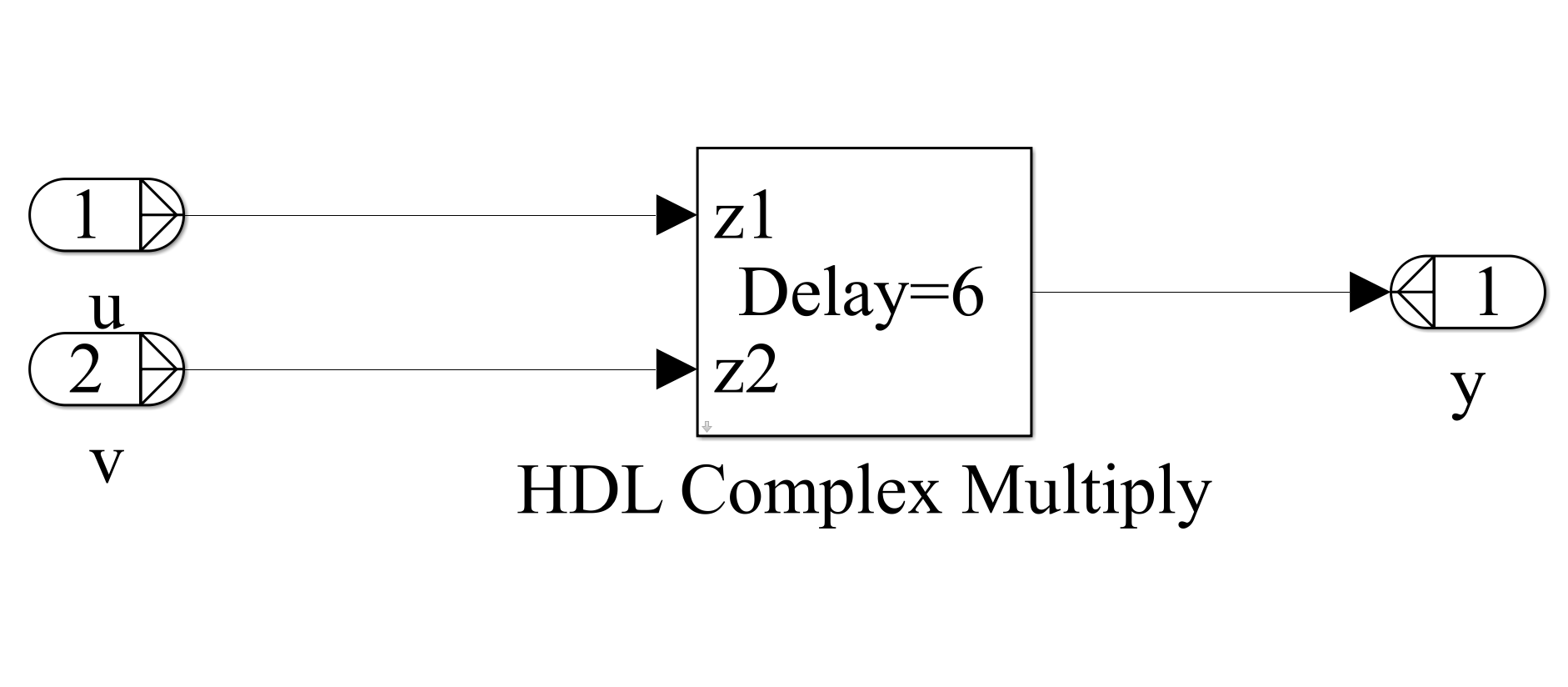} \\
(c)
\caption{(a) Hardware demonstration of the inner product in the QS-SVM-based DoA estimation of the proposed beamformer using HDL. (b) QS-SVM inner product. (c) HDL complex multiplication.}
\label{fig:inner_product}
\end{figure*}



\begin{figure*}[ht]
\centering
\includegraphics[width=1.0\textwidth]{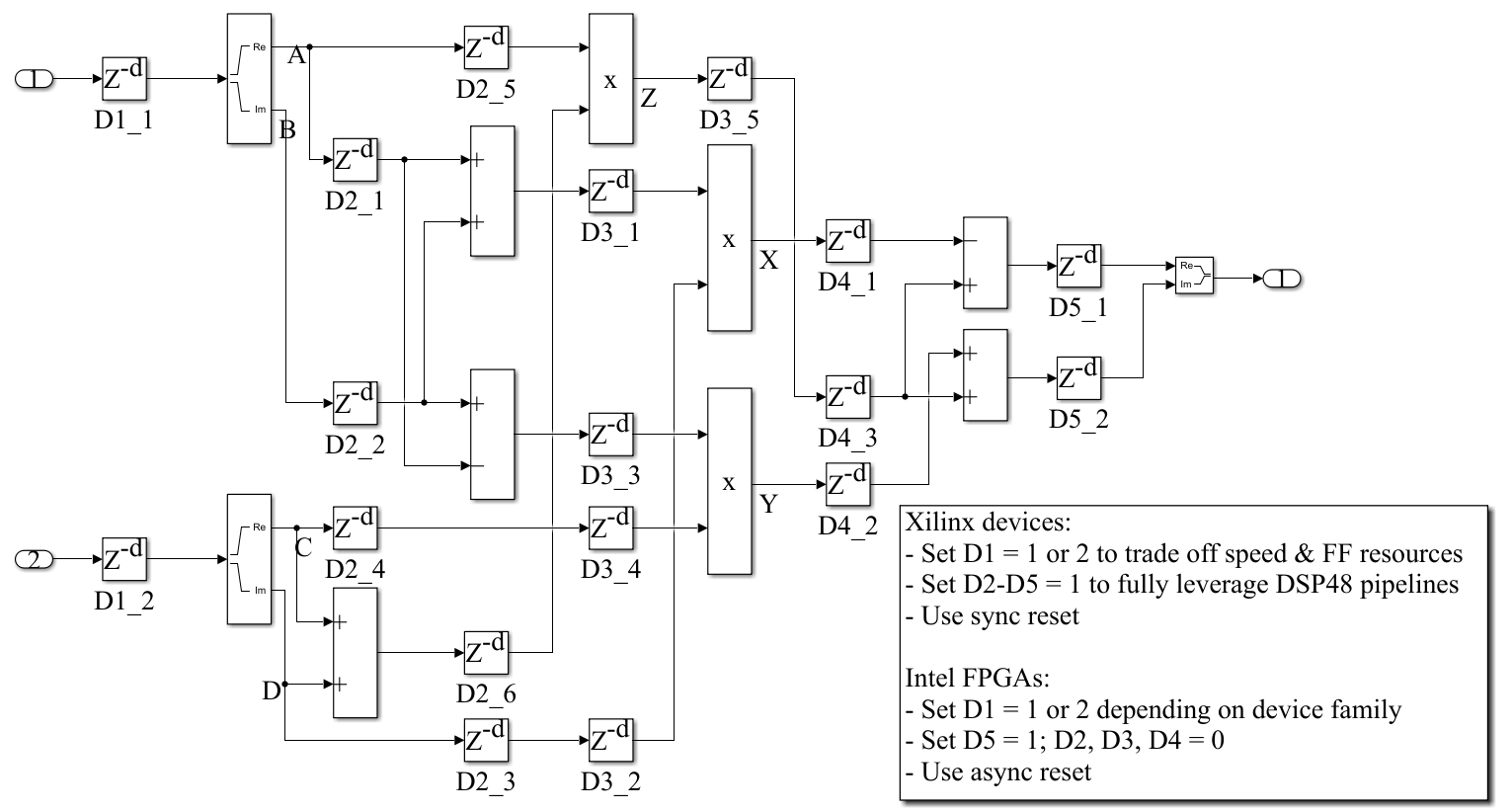}
\caption{The hardware implementation of HDL complex multiplications of the QS-SVM optimization method.}
\end{figure*}
The tree sum with distributed pipelining takes a sequence of tasks in which each sequence includes its own steps and overlaps with other steps with different tasks in the execution time. 
The practical implementation of synchronizing of the pipeline distributions can provide limitations, however performing a parallel merging on trees, as demonstrated in Fig. 9, can significantly overcome the aforementioned constraints~\cite{Tataria}. 
An increase in the number of pipeline stages results in a significant enhancement in the throughput of the FPGA and an increase in the operational clock frequency, however, an increase in the latency, which is not of practical interest. 
Therefore, we have to consider a trade-off between the performance enhancement in terms of the throughput of the FPGA and the operational clock frequency and latency by choosing an appropriate number of the pipeline stages. 
The proposed configuration of the pipelined inner product computations can drastically decrease delays in the mathematical operations, and power dissipation, and an area reduction.  

\subsection{Deliverable and Results of the QS-SVM-based Beamformer on the FPGA board}

In this section, the proposed architecture for the setup and implementation includes the FPGA board, as the hardware unit, integrated with MATLAB, as the processing unit. 
The proposed hybrid antenna array with bowtie elements consistent with Fig.~\ref{fig:configuration} has been rigorously employed for the QS-SVM-based digital beamformer.  
In Fig. 3, suppose only one of three available incident signals in the environment is desirable. 
In order to separate and then classify the three signals, we should primarily retrieve directions of all the three incident signals using the DoA estimation, described in detail in~\cite{Komeylian2021-2,Komeylian2021-3,HASHEMIFATH202080,Komeylian2020,Suykens,Darzi}. 
Consequently, the proposed digital beamformer has delivered the synthesized beam pattern, as illustrated in Fig. 11. 
The synthesized beam pattern consists of the two following tasks,

\begin{itemize}
    \item Null steering for undesired signals by replacing nulls of the radiation pattern of the proposed hybrid antenna array in the detected directions of arrival of undesired signals. Hence, we can weaken significantly or eliminate undesired signals. 

    \item Keeping the desired signal unchanged by exerting a power with a 0 dB level in the detected direction of arrival of the desired signal. We can neither strengthen nor weaken the desired signal, due to the two following reasons: (1) since the desired signal may include noise and other unwanted signals, thereby any amplification in the desired signal results in magnifying noise and other unwanted signals as well, and (2) any reduction in the desired signals is not our choice.
\end{itemize}

\begin{figure*}[ht]
\centering
\includegraphics[width=\textwidth]{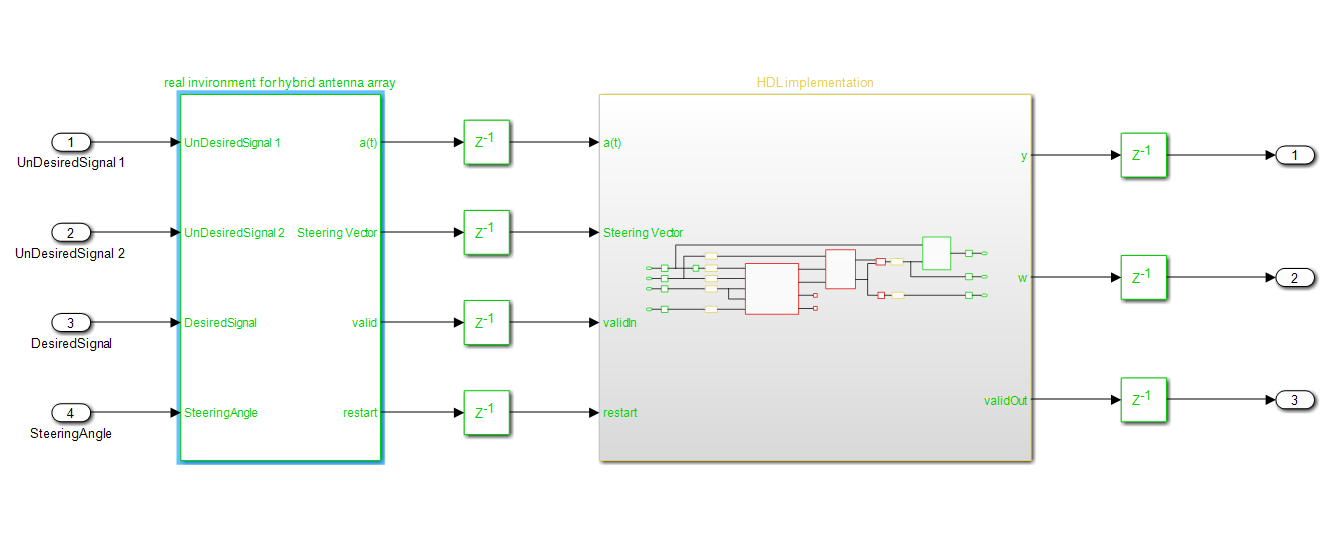}
\caption{The spatial filtering performance of the QS-SVM-based digital beamforer on the FPGA board. The obtained results are associated with the part of the pattern plot in Fig.~\ref{fig:doa_Estimator}.}
\label{fig:filtering}
\end{figure*}

\section{Performance Evaluation of the FPGA-based Beamformer} 
\label{sec:performance}

The discussion in the previous section focused on comprehensively demonstrating the deployment and setup of the digital beamformer. 
In the following section, we aim to evaluate and validate the digital beamformer performance. 
In this scenario, results-based monitoring allows for providing a framework to assess and evaluate the proposed digital beamformer performance in terms of (1) throughput evaluation, (2) average latency time, and (3) performance efficiency.   

\subsection{Throughput Evaluation}

The Euclidean distance refers to the most well-known distance metric for deep learning as well as the classification process in machine learning. 
The Euclidean distance metric allows for measuring the distance between paired examples in the high-dimensional feature space. 
However, the utilization of Euclidean distance results in neither preserving the correlation of pairs nor enhancing the robustness of pairs. 
To overcome the aforementioned limitations, Tongtong Yuan and his research group in~\cite{Yuan} have rigorously verified that the SNR distance is a more promising metric for measuring differences between pair features. 
Moreover, they extracted the SNR formulation for deep learning as the ratio of feature variance and noise variance. 
Since the throughput of the classification output is very much affected by SNR values, a drastic degradation in the SNR functionality causes a totally unacceptable throughput.
\begin{figure}[ht]
\centering
\includegraphics[width=0.6\textwidth]{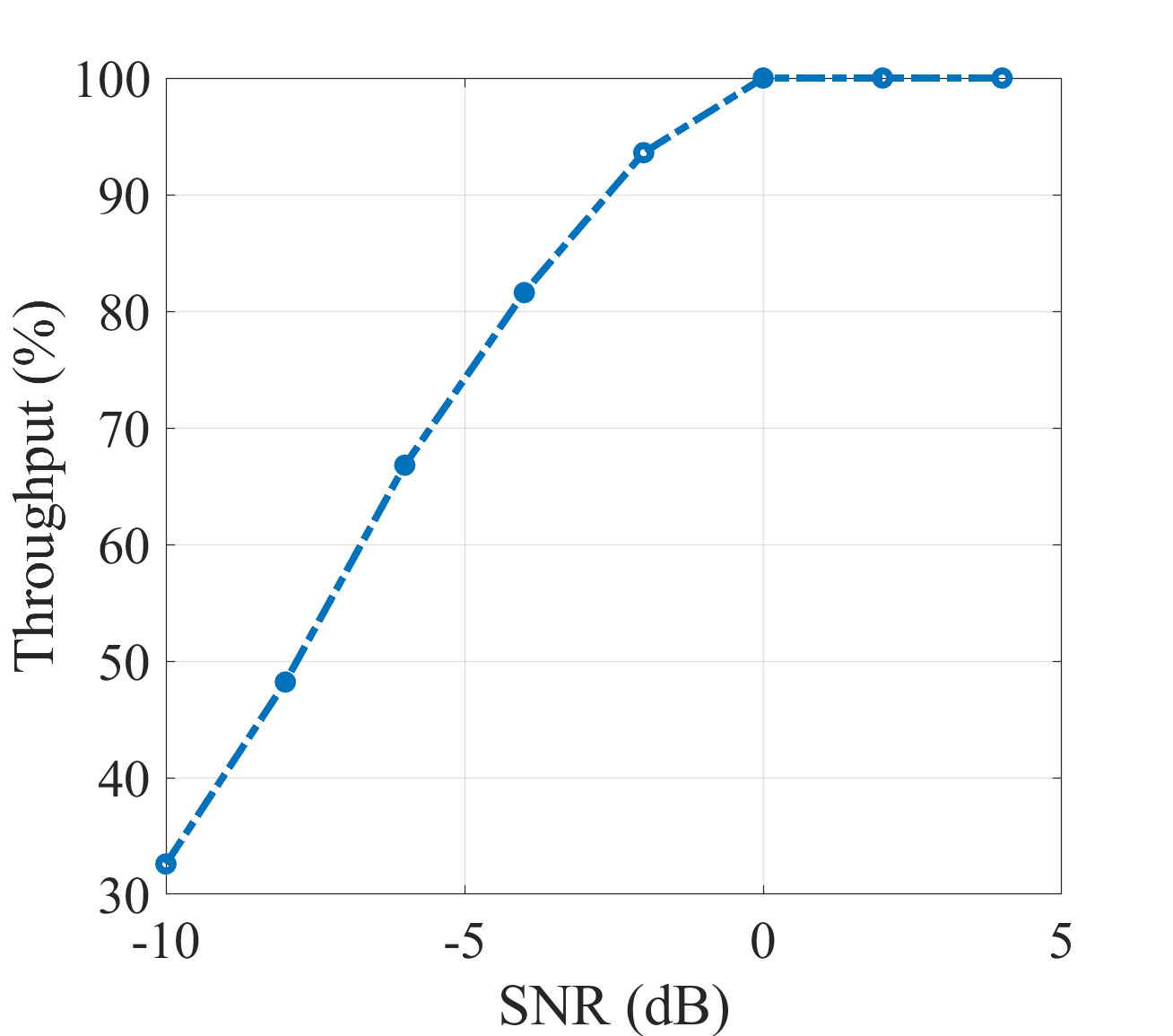}
\caption{Variation of throughputs of the classification performance of the QS-SVM-based beamformer in terms of SNRs consistent with Fig.~\ref{fig:doa_Estimator_a}.}
\label{fig:throughput}
\end{figure}

Fig.~\ref{fig:throughput} illustrates that the throughput of the classification output has significantly increased in the presence of greater SNR values. 
Similar results of Fig. 12 have been achieved in~\cite{Yuan}, in which the enlargement of inter-class distances for learned features resulted from the better performance of the SNR distance (or the higher throughput of the classification output) is accompanied by an increase of SNR values.  

\subsubsection{Latency Evaluation} 

\begin{figure}[ht]
\centering
\includegraphics[width=0.6\textwidth]{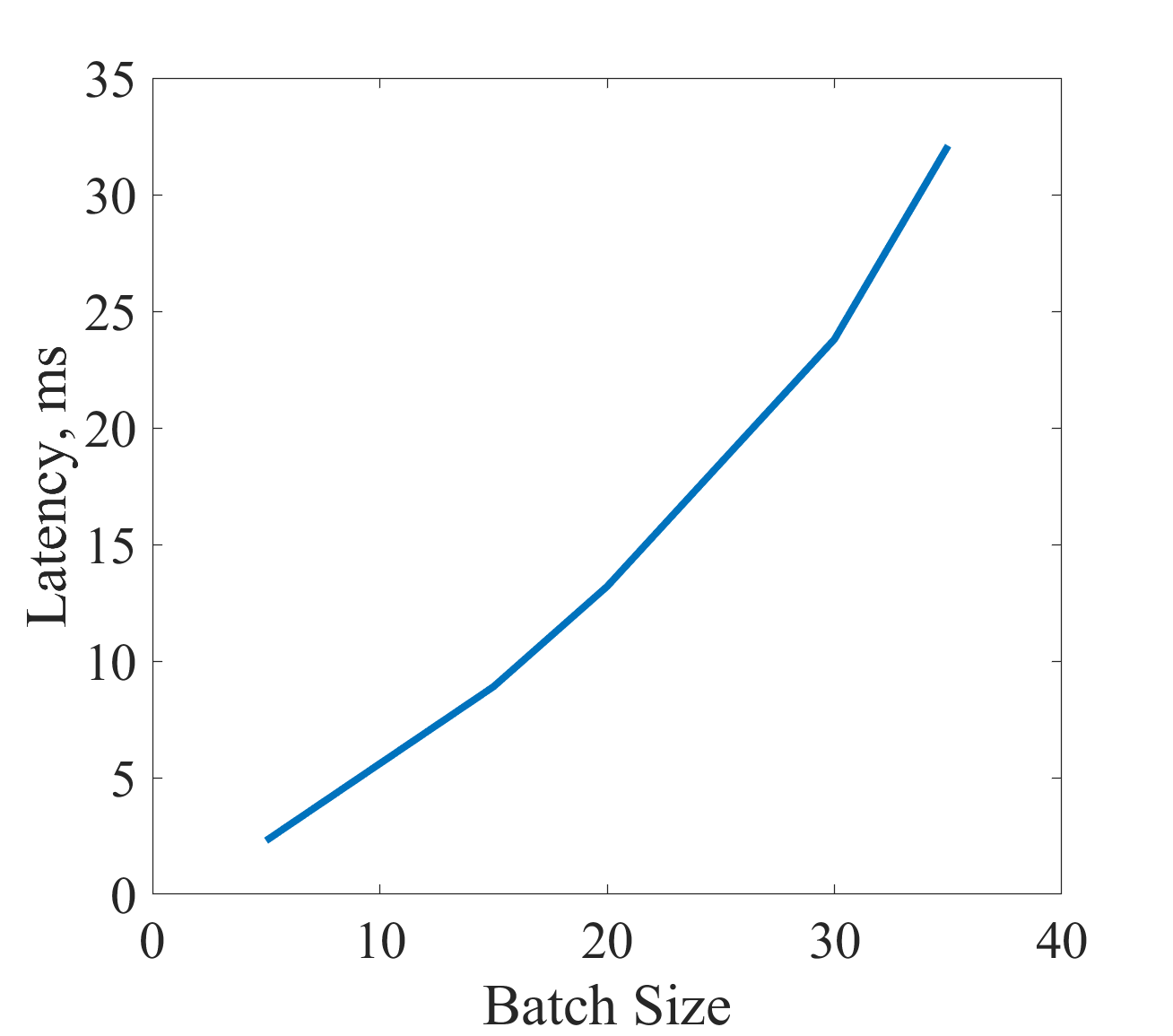}
\caption{Latency analysis for different batch sizes consistent with Fig.3. Latency is presented in milliseconds and averaged over the test set of the network.}
\label{fig:latency}
\end{figure}

Latency allows for measuring the average time for processing a fixed-point dataset in different sizes of a batch. 
In the supervised ML, a batch size refers to the length of training dataset in each iteration. 
In an ideal scenario, in which the FPGA integrated with a processor is working at speed, an average latency time is around or below 50 microseconds. 
Fig.~\ref{fig:latency} has reported the simulation results on the average latency time of the implementation setup of Fig.~\ref{fig:latency} in terms of batch sizes. 
Fig.~\ref{fig:latency} ensures that the latency time is very much affected by the length of the fixed-point training data. 
In other words, an increase in the length of the training dataset is accompanied by a reduction in the processing time. 

\subsubsection{Performance Efficiency} 

Fig.~\ref{fig:filtering} represents analogous results of the performance efficiency of the FPGA-based implementation, consistent with Fig.10, with the hybrid antenna array with the two different types of array elements. In this scenario, the efficiency of the ML-based DoA estimator for the hybrid antenna array with bowtie elements outperforms with its counterpart configuration with dipole elements more than 20\%.
\begin{table}
    \centering
    \bgroup
    \def\arraystretch{1.5}
    \begin{tabular}{P{4.0cm}|P{3.0cm}|P{3.0cm}} 
    & Hybrid antenna array with bowtie elements & Hybrid antenna array with dipole elements \\ \hline
    \pbox{20cm}{Performance 
    efficiency \\ of the QS-SVM beamformer} &  96\% & 75\%
    \end{tabular}
    \egroup
    \caption{A comparison between the classification performance of the QS-SVM beamformer using the hybrid antenna array with bowtie elements and dipole elements. The proposed hybrid antenna arrays have been demonstrated in~\cite{https://doi.org/10.48550/arxiv.2210.00317}.}
    \label{tab:performance}
\end{table}

\section{Conclusion}
\label{sec:conclusion}

To conclude, in this work, we have outlined and reviewed some of the recent techniques and proposals that employ beamformers for overcoming the limitations of stable and strong connectivities in a wireless communication channel. 

The kernel-free classifier of QS-SVM can effectively support non-linearly separable datasets without mapping them to any larger feature space in the DoA estimation. 
In addition, adding an extra regularizer to the objective function has enhanced the capability of the QS-SVM technique for linearly-separable datasets of the DoA estimation. 
MVDR beamforming technique is capable of updating weight vectors for the beamforming, but it has a weak performance in the nullsteering. 
We have demonstrated that the nullsteering performance has been strongly improved by performing an additional beamforming technique of LCMV, showing deep nulls with powers smaller than -10 dB in undesired signals. 

The configuration of the antenna array, which has been employed for the proposed beamformer, can significantly affect on the performance and functionality of the DoA estimation and beamforming techniques. 
The spatial distribution of the hybrid antenna array in the horizontal and vertical directions has further enhanced its capability in controlling reception signals. 
In the following, the software and hardware implementations of the QS-SVM-based beamformer have been fulfilled in the Mathworks' Simulink visual programming environment and FPGA board, respectively. 
Details of the hardware implementation have been fully demonstrated on the FPGA board. 

The proposed optimization techniques incorporated with the hybrid antenna array have resulted in a very strong beamformer with superior capabilities in nullsteering, a high throughput of about 100\%, a low average latency, and a high performance efficiency of more than 90\%. 




\ifCLASSOPTIONcaptionsoff
  \newpage
\fi

 
\bibliographystyle{IEEEtran}

\bibliography{IEEEabrv,bibtex/bib/antenna}







\end{document}